\documentclass[12pt,twoside,A4]{article}
\usepackage{thumbpdf,lmodern,bm}
\usepackage{graphicx,subcaption}
\usepackage{amsmath} 	        
\usepackage{amssymb}
\usepackage[table, dvipsnames]{xcolor}
\usepackage{apacite}
\usepackage{natbib}
\usepackage{fullpage}

\usepackage{hyperref}
\hypersetup{pdfstartview={XYZ null null 1.00}}

\begin{document}

\title{Generalized Linear Randomized Response Modeling using \textit{GLMMRR}}
\author{by Jean-Paul Fox \footnote{University of Twente, The Netherlands}, Konrad Klotzke\footnote{University of Twente, The Netherlands}, Duco Veen\footnote{University of Utrecht, The Netherlands}}

\maketitle

\abstract{
 Randomized response (RR) designs are used to collect response data about sensitive behaviors (e.g., criminal behavior, sexual desires). The modeling of RR data is more complex, since it requires a description of the RR process. For the class of generalized linear mixed models (GLMMs), the RR process can be represented by an adjusted  link function, which relates the expected RR to the linear predictor, for most common RR designs. The package \textit{GLMMRR} includes modified link functions for four different cumulative distributions (i.e., logistic, cumulative normal, gumbel, cauchy) for GLMs and GLMMs, where the package \textit{lme4} facilitates ML and REML estimation. The mixed modeling framework in \textit{GLMMRR} can be used to jointly analyse data collected under different designs (e.g., dual questioning, multilevel, mixed mode, repeated measurements designs, multiple-group designs). The well-known features of the GLM and GLMM (package \textit{lme4}) software are remained, while adding new model-fit tests, residual analyses, and plot functions to give support to a profound RR data analysis. Data of H\"{o}glinger and Jann (2018) and H\"{o}glinger, Jann, and Diekmann (2014) is used to illustrate the methodology and software.
}

\section[Introduction]{Introduction}
The randomized response technique (RRT) has been developed to encourage respondents to answer truthfully to questions about sensitive behaviors. The RRT is designed to collect responses in an indirect way, where respondents are instructed to answer a sensitive question truthfully only with a certain probability. Therefore, an affirmative response to a sensitive question is masked by a random device at the individual level. The traditional way of direct questioning (DQ) about sensitive topics is known to lead to socially desirable responses or non-cooperation of the respondents. Honest responding is encouraged by RRT by increasing anonymity and confidentiality.

After the introduction of the randomized response (RR) design by \citet{warner1965randomized}, many randomized response (RR) designs have been introduced to improve RRT performance in collecting sensitive information. Warner's design is based on a deck of cards containing cards with the sensitive question and cards with its negation. Since then, different randomizing devices have been introduced to improve the privacy protection of the respondent. Outcomes of rolling a dice or spinning a spinner have been used to determine at random one of the possible routings to a response for a respondent. Another non-sensitive question has been introduced to make the design easier to implement, where the response to the non-sensitive question is used to steer the response process at random. For online surveys, an online randomizing device such as a digital dice or spinner is less reliable, since the computer outcomes can be stored. Therefore, techniques have been developed where a user-interaction is requested to generate a random outcome, which cannot be digitally stored. Currently, there are many ways to randomize the response in an online survey, where the randomizing design properties -- the level of privacy protection and the efficiency of the design -- are still under the control of the investigator.

The flexibility in RR designs has also improved the utility of the RRT. They can be applied to survey items \citep{lensvelt2005meta,hout2010estimating} but also to scale items for measuring sensitive constructs \citep{fox2005multilevel, fox2008using, fox2013mixture}. Different RRT implementations have been introduced and applied in many research fields. Together with this expansion of the RRT to different domains, statistical tools have been developed to improve the analysis of RR data. \citet{scheers1988covariate} developed a (logistic) linear regression method for RR data to explore relationships between variables in addition to estimating prevalence rates of the sensitive attribute. Since then, more advanced statistical regression models for RR data have been developed. For instance, randomized item-response theory (RIRT) models to measure sensitive constructs \citep{fox2016bayesian}, and mixture RR models to deal with non-cooperating participants \citep{fox2013mixture}. To further stimulate the use of RRT, software tools are needed to make the developed methods and designs for RR data widely available.

Recently, two packages for RR data have become available in R. The package \textit{rr} of \citet{blair2015rr} is developed for a univariate power analysis to measure the sensitive item prevalence for four RR designs. It also comprehends a logistic regression function for RR data collected with a single RR design. The package \textit{RRreg} of \citet{JSSv085i02} has logistic and linear regression routines, which can include random effects, and tools for power analysis for different RR designs (including those for continuous data). However, both packages are restricted to single-group RR designs and do not allow a joint analysis of different randomization schemes across items and/or participants.

Multiple-group RR designs are particularly interesting in online surveys, where different RR designs and different random devices can be easily varied. For instance, the level of protection can be moderated by varying the design parameters, and RR designs can be varied across survey items to improve truth-telling. For validation studies, multiple-group RR designs are relevant to make comparisons between groups questioned with different RR designs \citep{hoglinger2018more, hoglinger2016sensitive}. For a meta-analysis, a joint analysis of the available data can improve the measurement of a sensitive prevalence. Furthermore, differences in prevalence estimates can be examined across studies. More recently, there is an increased interest in measuring the efficacy of the RRT to improve honest disclosure of sensitive information \citep{john2018and}. This is usually done through validation studies where a DQ group serves as the baseline, and it is investigated whether higher prevalence estimates are obtained with a single-group RRT. The huge amount of literature on the topic, and mixed results about the benefits of RRT, has led to skeptical positions toward RRT \citep{wolter2013asking}. However, the RR design parameters (the random-device properties) as well as the type of RR design can influence the response behavior, and care must be taken in designing the RRT to reduce the level of misreporting. It is not likely that a fixed privacy level encourages truth-telling in a uniform way across participants. The performance of RRT will vary across subgroups and participants, and different RR designs can vary in performance across subgroups depending on the level of sensitivity of the question.

The R package \textit{GLMMRR} extends existing implementations by providing generalized regression tools for multiple-group RR designs. It builds in a natural way on common regression methods of the system package \textit{stats} to make them suitable for RR data. The R package \textit{GLMMRR} gives supports to \textit{generalized linear (mixed effects) regression modeling} of binary RR data and extends the popular generalized linear regression routines in R to handle RR outcomes, by including RR-specific link functions for a wide variety of RR designs. The \textit{GLMMRR} provides several important contributions for the joint regression analysis of binary RR data:
\begin{enumerate}
\item Different link functions (logit, probit, complementary log-log, cauchit) are supported to optimize the link between the linear predictor and the response and to avoid restrictions on the range of the (expected) response. The complementary log-log and cauchit link functions approach the asymptotes of zero and one asymmetrically.
\item Each link function is modified to an RR-link function to make it suitable for various RR designs (Warner, Unrelated question, Forced response, Kuk, Crosswise, Triangular), where two design parameters specify the properties of the RR design.
\item The package supports a joint regression analysis of RR data sampled with different RR designs and different design parameters. Each response observation is either indirectly observed with an RR design with unique design parameters or directly observed without an RR design.
\item For randomized item-response data, RR-design specific (weighted) prevalence rates and confidence intervals can be computed for each item. Trait levels can be estimated, where the items serve as indicators of a sensitive trait.
\item Extensive residual tools are available for the evaluation of the fit of the model. This includes Pearson, deviance and response residuals, goodness-of-fit statistics (Pearson's and deviance chi-square statistic, Hosmer-Lemeshow statistic), and the Akaike's Information Criterion (AIC) and the Bayesian Information Criterion (BIC).
\item The \textbf{RRglm} is used to fit a GLM and an object of class \textbf{"RRglm"} is created, which extends the class \textbf{"glm"} with RR data. The \textbf{glm} methods are applicable to an object of class \textbf{"RRglm"}. The function \textbf{summary} or \textbf{print} can be used to obtain a summary of the results. Other useful features can be extracted using the well-known generic functions, such as \textbf{coefficients}, \textbf{effects}, \textbf{fitted.values} and \textbf{anova}.
\item The \textbf{RRglmer} for fitting a GLMM returns an object of class \textbf{"RRglmerMod"} which extends the class \textbf{"glmerMod"} and \textbf{"merMod"} with RR data. The methods for an object of class \textbf{"glmerMod"} are applicable for an object of class \textbf{"RRglmerMod"} (the methods can be found in the \textit{lme4} documentation).
\end{enumerate}

 This paper is organized as follows. In the next Section, the RR-link functions are introduced, and the GLM and GLMM modeling framework for RR data are introduced. A general introduction is given where the responses are distributed according to a distribution of the exponential family, and the RR-link function is used to link the linear predictor to the expected (randomized) response. In Section \ref{MLestimation}, the ML and REML estimation method is discussed for the GLM(M) with RR-link functions. Then, the RIRT modeling approach for measuring sensitive constructs is discussed. Section \ref{modelfit} discusses different methods to evaluate the fit of the model.  Different residuals are defined (Pearson, deviance, response) for a GLM. For a GLMM, a conditional residual is defined as the difference between the observation and the conditional expected value (given the random effect estimate). Goodness-of-fit statistics are introduced by grouping the Pearson and deviance residuals. For a fixed number of groups, the Hosmer-Lemeshow (H-L), the Pearson and the deviance statistic are approximately chi-square distributed, and can be used to evaluate the global fit of a GLM. In Section \ref{software}, the two main functions of the package are discussed for fitting GLMs and GLMMs with RR data. In Section \ref{applications}, an illustration of the package is shown by analysing data from the validation study of \citet{hoglinger2018more}, where differences in outcomes between RR designs are evaluated. Furthermore, the prevalence of student misconduct is analysed with a randomized item-response theory model where students were assigned to different treatment groups, each assigned to a specific RR design. A joint modeling approach is carried out to examine differences between treatment groups. Then, in Section \ref{conclusion}, the conclusions are given.

\section[General randomized-response probability model]{General randomized-response probability model}\label{RRtechniques}

When collecting RR data, each observation is randomized before it is observed. The RR process makes it impossible to directly relate an observed RR to the sensitive question. It masks the answers of respondents. To ease the notation, a distinction is made between respondents with index $i$ and items with index $k$. A further specification is possible where participants and/or items are grouped, which will be shown in the real data examples in Section \ref{applications}. Thus, respondent's $i$ prevalence to the sensitive question $k$, denoted as $\tilde{\pi}_{ik}$, cannot be directly measured. Instead, the RR probability, denoted as $\pi_{ik}$, is measured.

\citet{fox2018generalized} showed that for the common RR designs for binary data, the RR probability for the observed RR data can be related to the prevalence of the sensitive question through a linear equation. This linear relationship is given by,
\begin{eqnarray}
\pi_{ik} & = & c_{ik} + d_{ik} \tilde{\pi}_{ik}. \label{RRmodel}
\end{eqnarray}
The RR parameters $c$ and $d$ determine the type of RR design. A variety of RR designs can be represented by Equation (\ref{RRmodel}): Warner's design \citep{warner1965randomized}, Unrelated Question (UQM) design, Forced response (FR) design \citep{boruch1971maintaining}, Kuk's design \citep{kuk1990asking}, Triangular design \citep{yu2008two}, and the Crosswise (CW) design \citep{yu2008two}. The RR parameters $c$ and $d$ are retrieved from the RR design parameters. For instance, for Warner's design, a positive response is given to the sensitive question with probability $p_1$ or a negative response to its negation with probability $1-p_1$,
\begin{eqnarray}
\pi_{ik} & = & p_1\tilde{\pi}_{ik} + (1-p_1)(1-\tilde{\pi}_{ik}) \nonumber \\
& = & \underbrace{(1-p_1)}_{c_{ik}} + \underbrace{(2p_1-1)}_{d_{ik}}\tilde{\pi}_{ik}. \nonumber
\end{eqnarray}
 The parameters $c_{ik}$ and $d_{ik}$ describe the random response process, which are allowed to vary across questions and respondents such that the data-collection design is defined for each single response. \citet{fox2018generalized} gives an overview in which the RR-design parameters are represented as RR parameters for various RRTs.

Although the RR designs work in different ways, a general RR model can be defined, which relates the RR data to the prevalence of the sensitive question. This general RR model is extended to a GLM and GLMM by linking the prevalence $\tilde{\pi}_{ik}$ to a linear predictor. The \textit{GLMMRR} functions require the RR design parameters $p_1$ and $p_2$. The function \textbf{getRRparameters} provides the RR parameters $c$ and $d$ given the type of RR design and corresponding parameter values. This is illustrated in our applications in Section \ref{applications}, where the RR designs UQM, FR, and CW are used and the DQ design serves as a baseline.

\subsection[Modified link functions for GLMs and GLMMs]{Modified link functions for GLMs and GLMMs}\label{modifiedlinks0}
The object is to define the natural link function which relates the expected RR to a linear predictor for the prevalence $\tilde{\pi}_{ik}$. The RR parameters  $c_{ik}$ and $d_{ik}$ are integrated in the link function to relate the expected RR observation to this linear predictor without explicitly parameterizing the prevalence. In this approach, the natural link functions for Bernoulli distributed observations are modified to account for the RR parameters. Simply by adjusting the natural link functions, the class of GLMs and GLMMs is extended to (binary) RR data.

Let $Y_{ik}$ denote the (binary) RR observation of respondent $i$ to item $k$. For binary data, the expected response is equal to the RR probability, $\pi_{ik}$. The expected response is related to a linear predictor denoted as $\eta_{ik}$, which is a linear combination of predictor variables for respondent $i$ related to the response to item $k$. For binary RR data, the expected RR observation is equal to the RR probability, $\pi_{ik}$, and the relationship with the $\eta_{ik}$ can be represented by
\begin{eqnarray}
E\left(Y_{ik} \mid \eta_{ik}\right)  & = & \pi_{ik} \nonumber \\
 & = & c_{ik} + d_{ik} \, \tilde{\pi}_{ik}  \nonumber \\
 & = & c_{ik} + d_{ik} g^{-1}\left( \eta_{ik} \right) \nonumber \\
 & = & c_{ik} + d_{ik} g^{-1}\left(\mathbf{x}^t_{ik}\bm{\beta} + \mathbf{z}^t_{ik}\bm{b}_{i}\right), \label{GRModel}
\end{eqnarray}
 where the explanatory variables $\mathbf{x}_{ik}$ have fixed effects (i.e., common across respondents) and the $\mathbf{z}^t_{ik}$ have random effects (i.e., vary across respondents). The random effect $\bm{b}_{i}$ is assumed to have a multivariate normal distribution with mean zero and variance $\bm{\Sigma_b}$. For $c_{ik}=0$ and $d_{ik}=1$, the $g^{-1}()$ is the mean function for (non-randomized) Bernoulli distributed observations. It follows that the natural link function for Bernoulli distributed data, $g()$, needs to be modified to link the expected RR to the linear predictor for the prevalence $\tilde{\pi}_{ik}$,
 \begin{eqnarray}
g\left(\frac{\pi_{ik} - c_{ik}}{d_{ik}} \right) & = & \eta_{ik}.
\end{eqnarray}
Consider the Bernoulli distributed RR observation with success probability $\pi_{ik}$, for which four different cumulative distribution functions can be applied (i.e., logistic, probit, gumbel, cauchy). Then, the distribution of the RR is given by
\begin{eqnarray}
Y_{ik} &\sim& \mathcal{B}\left(\pi_{ik} \right) \\
\pi_{ik} & = & c_{ik} + d_{ik} g^{-1} \left(\eta_{ik}\right) \\
& = &
\left\{
  \begin{array}{ll}
   c_{ik} + d_{ik}\frac{\exp\left(\eta_{ik}\right)}{1+\exp\left(\eta_{ik}\right)} & \textrm{Logistic} \\
   c_{ik} + d_{ik}\Phi\left(\eta_{ik} \right)  & \textrm{Probit} \\
   c_{ik} + d_{ik}\left(1- \exp\left(1-\exp\left(\eta_{ik}\right)\right)\right) & \textrm{Gumbel} \\
   c_{ik} + d_{ik}\left(\arctan\left(\eta_{ik}\right)/\pi + \frac{1}{2}\right)  & \textrm{Cauchy}. \\
  \end{array}
\right.\label{naturaldistributions}
\end{eqnarray}
Next, four different link functions are defined by modifying the linear predictor's relation with the prevalence, such that it relates to the expected RR observation. The possible (modified) link functions are given by,
\begin{eqnarray}
\eta_{ik} & = & g\left((\pi_{ik}-c_{ik})/d_{ik} \right) \\
& = &
\left\{
  \begin{array}{ll}
  \ln\left(\frac{\pi_{ik}-c_{ik}}{c_{ik}+d_{ik} - \pi_{ik}} \right) & \textrm{Logit Link} \\
   \Phi^{-1}\left(\frac{\pi_{ik}-c_{ik}}{d_{ik}} \right)  & \textrm{Probit Link} \\
   \ln\left(-\ln\left(\frac{c_{ik} + d_{ik} - \pi_{ik}}{d_{ik}} \right)\right)  & \textrm{Complementary log-log Link} \\
   \tan\left(\pi\left(\frac{\pi_{ik}-c_{ik}}{d_{ik}}  \right)\right)  & \textrm{Cauchit Link}. \\
  \end{array}
\right. \label{modifiedlinks}
\end{eqnarray}
The considered four different link functions are modified versions of the common link functions as defined in, for example, \citet{mccullagh1989generalized} and \citet{tutz2011regression}. When $c_{ik}=0$ and $d_{ik}=1$, the common link functions for directly observed responses are given.

\subsubsection{Exponential family distributions}
The modified (natural) link functions can be employed for (Bernoulli) exponential family distributed RR data. Assume the observed RR data are distributed according to a distribution of the exponential family; that is,
\begin{eqnarray}
p\left(y_{ik} \mid \theta_{ik}, \phi_{ik} \right) & = & \exp\left(\frac{ {y}_{ik} \theta_{ik} - A\left(\theta_{ik} \right)}{\phi} + C\left( {y}_{ik},\phi \right) \right).
\end{eqnarray}
Then, the log-likelihood of the parameter $\theta_{ik}$ and $\phi$ is expressed as
\begin{eqnarray}
l\left(\theta_{ik}, \phi  ;  {\mathbf{y}}  \right) & = & \log p\left( {\mathbf{y}}  \mid \theta_{ik}, \phi \right) = \sum_{i,k} \frac{y_{ik} \theta_{ik} - A\left(\theta_{ik} \right)}{\phi} + C\left(  {y}_{ik},\phi \right).
\end{eqnarray}
The $\theta_{ik}$ is the canonical parameter and depends via a linear predictor on explanatory variables. The dispersion parameter is usually unknown and used to model the variance of the response data. The functions $A(.)$ and $C(.)$ are known and determined by the specified distribution of the family. Then, for Bernoulli distributed RR observations the natural form of the parameters is given by
\begin{eqnarray}
\theta_{ik} &=& \log\left(\frac{c_{ik} + d_{ik}\tilde{\pi}_{ik}}{1-(c_{ik} + d_{ik}\tilde{\pi}_{ik})}\right) \\
A(\theta_{ik}) &=& \log\left( 1+\exp\left(\theta_{ik}\right)\right) \\
\phi &=& 1.
\end{eqnarray}

Thus, the general properties of the exponential family distributions can be used to make inferences about the parameters. For instance, the maximum likelihood estimate for the population prevalence $\tilde{\pi}$ can be derived. The log-likelihood has a unique maximum at $\hat{\theta}$ which is the solution to
\begin{eqnarray}
\sum_{i,k} y_{ik} & = & A'(\hat{\theta})  =  \sum_{i,k} \left(1+\exp(-\hat{\theta}_{ik})\right)^{-1}  \nonumber \\
\sum_{i,k} y_{ik} & = & \sum_{i,k} \left(c_{ik}+d_{ik} \hat{\pi}\right) \nonumber \\
\hat{\pi} & = & \left(\sum_{i,k} y_{ik} - \sum_{i,k} c_{ik}\right)/ \sum_{i,k} d_{ik} = (\overline{y} - \overline{c})/\overline{d}. \label{MLestimate}
\end{eqnarray}
This maximum likelihood (ML) estimate for the prevalence is referred to as a weighted (ML) estimate, where the weights are defined by the RR parameters.

Furthermore, the inverse of the function $A'(\theta)$ represents the natural link function. Thus, the modified natural link function for Bernoulli distributed RR data can be derived using this property of the exponential family distribution:
\begin{eqnarray}
\pi_{ik} &=& A'\left(\theta_{ik}\right) \nonumber \\
& = & \left[1+\exp\left(-\log\left(\frac{c_{ik}+d_{ik}g^{-1}(\eta_{ik})}{1-(c_{ik}+d_{ik}g^{-1}(\eta_{ik}))}\right) \right) \right]^{-1} \nonumber \\
& = & c_{ik}+d_{ik}g^{-1}(\eta_{ik}) \nonumber \\
\eta_{ik} & = & g\left((\pi_{ik} - c_{ik})/d_{ik}\right), \nonumber
\end{eqnarray}
where $g^{-1}(.)$ is the cumulative distribution function for (non-randomized) Bernoulli distributed data as defined in Equation (\ref{naturaldistributions}). The corresponding link functions are defined in Equation (\ref{modifiedlinks}).

\section{ML and REML estimation}\label{MLestimation}
It can be shown that the maximum likelihood (ML) equations for the GLM for RR data resemble the general form of the GLM ML-equations \citep[Appendix~B]{fox2018generalized}. The only difference is that the conditional expected RR observation includes the RR parameters and a modified link function is needed to link the conditional expected response to the linear term. The GLM parameters are usually estimated by ML methods using the iterative weighted least squares (IWLS) algorithm or Fisher scoring algorithm. The \textbf{glm} function implemented in R is a very flexible implementation of the general GLM framework \citep{chambers1992statistical}. The package \textit{GLMMRR} provides an expanded version of this function (\textbf{RRglm}), which includes modified link functions to fit GLMs on RR data. Given ML estimates, (maximum) likelihood theory can be used to obtain likelihood ratio tests, Wald and score tests.

The likelihood equations for the fixed effect parameters of the GLMM for RR data have the same structure as those for the GLMM \citep[Appendix~B]{fox2018generalized}. For the fixed effects, the GLM methodology can be used, since the fixed effects are not included in the random effect distribution. Depending on the dimension of the random effect parameter, numerical approximations are required to approximate the integrals to estimate the variance components and random effects. The numerical methods available in \textit{GLMMRR} build on those available in the package \textit{lme4}. Different numerical methods have been proposed; Laplace approximation and adaptive Gaussian quadrature are both implemented. Laplace approximation is usually fast and the default. The approximation improves when the cluster sizes increase. In Gaussian quadrature a number of quadrature points need to be chosen, and the approximation is improved by increasing the number of quadrature points. Adaptive Gaussian quadrature usually fails, when the dimension of the random effects is larger than two. It is also possible to compute restricted ML estimates (REML), which is also implemented. The \textbf{control} argument can be used in the function call to \textbf{RRglmer} to set the control parameters, which includes the optimizer to be used.

\section[Randomized item-response theory modeling]{Randomized item-response theory modeling}
An important class of item response theory (IRT) models belong to the GLMM class \citep{rijmen2003nonlinear}. These IRT models have a linear component for the transformed expected values of a binary response variable, which contains a random component(s) representing the latent variable(s). In the most common form, the linear term has a random component representing  a random person effect and fixed components representing item effects. In Section \ref{modifiedlinks0}, it is shown that the GLMM is extended to RR data using modified link functions. Therefore, IRT models belonging to the GLMM class can also be generalized to randomized item-response theory (RIRT) models.

The randomized item-response observations are clustered by persons and items, with $c_{ik}$ and $d_{ik}$ representing the RR parameters. The latent variable for person $i$ is denoted by $\vartheta_{i}$ and the effect of item $k$ by $\beta_k$. Then, the RIRT model -- the Rasch model for RR data -- can be represented as
\begin{eqnarray}
\pi_{ik} & = & c_{ik} + d_{ik}g^{-1}\left(\vartheta_i + \beta_k \right) \label{RIRTmodel} \\
\vartheta_i & \sim & N(0,\sigma^2) \nonumber
\end{eqnarray}
where $g^{-1}$ is usually the logistic or cumulative normal distribution function. The item effects have a positive sign and should be interpreted as a easiness parameters.

The RIRT model can be fitted in R using the function \textbf{RRglmer} from our package \textit{GLMMRR}. This is similar to fitting an IRT model using the function \textbf{glmer} from \textit{lme4} \citep{de2011estimation}, except that a modified link function is required for the RIRT. The data needs to be in long format. Then, each data case has a response (\textbf{response}), a person identifier (\textbf{person}), an item identifier (\textbf{item}), the type of RR design (\textbf{RRmodel}), and RR design parameters (\textbf{RRp1, RRp2}). The RR model and parameters are allowed to vary across data cases.

To fit the RIRT model in Equation (\ref{RIRTmodel}), a linear component is defined from the factor variables: \textbf{-1+item+(1|person)}. The \textbf{-1} restricts the general mean to zero, the \textbf{item} represents the item parameters, and the \textbf{(1|person)} represents the latent variable (random effect). This linear component is set equal to the outcome \textbf{response} in the model formula. To fit a logistic RIRT, the call to \textbf{RRglmer} includes a modified logistic link function (\textbf{RRlink.logit}):
 \begin{verbatim}
RRglmer(response ~ -1 + item + (1|person), link = "RRlink.logit",
    RRmodel = RRmodel, p1=RRp1, p2=RRp2, data=data)
\end{verbatim}
where the \textbf{RRmodel}, \textbf{p1} and \textbf{p2} arguments define the RR design for each single response. The probit RIRT model can be fitted using the link function \textbf{RRlink.probit}. The RIRT models can be extended by including (1) item-covariate models (e.g. linear-logistic test model), (2) person-covariate models (e.g. multilevel IRT), and (3) person-by-item covariate models. Covariates with fixed or random effects can be included in the linear term.

\section{Model fit and diagnostics}\label{modelfit}

The likelihood ratio (LR) test can be used to compare nested models. The nested model is a restriction of a more general model by restricting one or more parameters most often to zero. The log-likelihood ratio (multiplied by minus two) is asymptotically chi-square distributed with the degrees of freedom equal to the difference in the number of free parameters. The LR test can be performed with the \textbf{anova} function to compare two nested GLM(M)s for RR data. Note that the LR test cannot be applied to test a hypothesis on the boundary of the parameter space, since then the LR statistic is no longer chi-square distributed. ML estimation is preferred, when comparing models that only differ in their fixed part.

To compare (non)-nested models the usual information criteria can be used. The AIC and the BIC are both computed for GLM and GLMM, with the function \textbf{RRglm} and \textbf{RRglmer}, respectively, and reported in their output. The \textbf{anova} function also reports the AIC and BIC.

To evaluate the significance of fixed effects a z-statistic is reported, which is the ratio of the parameter estimate and the estimated standard error. The z-statistic is asymptotically equivalent to the LR test. P-values are reported in the output of \textbf{RRglm} and \textbf{RRglmer} with the assumption that the z-statistic is asymptotically normally distributed.

\subsection{Residuals}
The error term in GLM(M)s represents a Bernoulli random error term, and the errors are assumed to be independently distributed. There are different types of residuals and different types of residual sum of squares to examine the fit of the model. The estimated residuals in the GLM and GLMM are based on the fitted RR probabilities, $\hat{\pi}_{ik}$. The fitted prevalence can be computed as $\hat{\tilde{\pi}}_{ik} = (\hat{\pi}_{ik}-c_{ik})/d_{ik}$, according to Equation (\ref{RRmodel}).

\paragraph{GLM}
The residuals are computed for each observation $y_{ik}$, but for notational convenience the index $k$ is dropped, and the index $i$ refers to a single observation $y_{i}$. The Pearson and deviance residual is often computed and both are used in a goodness-of-fit statistic. The Pearson residual is defined as the standardized difference of the response and its expected value,
\begin{eqnarray}
r_p(y_i,\hat{\pi}_i) & = & \frac{y_{i} - \hat{\pi}_{i}}{\sqrt{\hat{\pi}_{i}(1-\hat{\pi}_{i})}}. \nonumber
\end{eqnarray}
The covariate patterns can be used to group the residuals in a natural way. Each unique covariate pattern defines a cluster $j$, where the number of response observations in cluster $j$ is given by $n_j$, the observed proportion $\overline{y}$, and a unique estimated probability denoted as $\hat{\pi}_j$. Then, the Pearson residual for cluster $j$ is defined as
\begin{eqnarray}
r_p(\overline{y}_j,\hat{\pi}_j) & = & \frac{\overline{y}_{j} - \hat{\pi}_{j}}{\sqrt{\hat{\pi}_{j}(1-\hat{\pi}_{j})/n_j}}. \nonumber
\end{eqnarray}
The clustering of observations can also be based on the predicted outcome \citep{hosmer1980goodness}. The observations are sorted according to their fitted probabilities, and this sorted vector is divided into $J$ clusters of equal size. The corresponding clustered Pearson residuals are referred to as H-L residuals. The Pearson residuals and clustered Pearson and H-L residuals can be extracted from a fitted object of class \textbf{RRglm} with the following commands, respectively:
\begin{verbatim}
residuals(object, type="pearson")
residuals(object, type="pearson.grouped")
residuals(object, type="hosmer-lemeshow", ngroups=10)
\end{verbatim}
For the H-L residuals,  the \textbf{ngroups} argument defines the desired number of clusters and the default is 10. The sum of squared grouped residuals defines a goodness-of-fit statistic, which is given by
\begin{eqnarray}
X^2 & = & \sum_{j=1}^{J} r^2(\overline{y}_j,\hat{\pi}_j)  =  \sum_{j=1}^{J} \frac{n_j \left(\overline{y}_{j} - \hat{\pi}_{j}\right)^2}{\hat{\pi}_{j}(1-\hat{\pi}_{j})}. \nonumber
\end{eqnarray}
Depending on the type of clustering, the statistic is the Pearson goodness-of-fit statistic, $X_p^2$, or the H-L goodness-of-fit statistic, $X_{HL}^2$. The Pearson statistic is asymptotically chi-square distributed for a fixed number of groups with degrees of freedom $J-(q+1)$, and $q$ the number of independent covariates. When the predictors variables are continuous the $X_p^2$ test cannot be used. For semi-continuous observations it is possible to find a clustering with a sufficient number of observations in each cluster. The $X_{HL}^2$ statistic is assumed to be chi-square distributed with $J-2$ degrees of freedom \citep{hosmer1980goodness}. The $X_p^2$ and $X_{HL}^2$ statistics can be computed from a fitted object of class \textbf{RRglm} using the function \textbf{RRglmGOF}:
\begin{verbatim}
RRglmGOF(object, doPearson = TRUE, doHlemeshow = TRUE,
    hlemeshowGroups = 10, rm.na = TRUE)
\end{verbatim}
The default number of clusters for the $X_{HL}^2$ statistic is ten, and data cases with missing observations are excluded. The deviance residual is defined as,
\begin{eqnarray}
r_{D}(\bar{y}_{i},\hat{\pi}_{i}) = sign(\bar{y}_{i} - \hat{\pi}_{i}) \sqrt{2 n_{i} \left(\bar{y}_{i}log\left(\frac{\bar{y}_{i}}{\hat{\pi}_{i}}\right) +
    (1 - \bar{y}_{i})log\left(\frac{1 - \bar{y}_{i}}{1 - \hat{\pi}_{i}}\right)\right)}
\end{eqnarray}
where $sign(\bar{y}_{i} - \hat{\pi}_{i})=1$ when $\bar{y}_{i} \geq \hat{\pi}_{i}$ and -1 when $\bar{y}_{i} < \hat{\pi}_{i}$. The grouped deviance residual is defined for cluster $i$ with $n_i>1$ and denoted as $r_d(\overline{y}_i,\hat{\pi}_i)$. For clustered observations, the sum of squared deviance residuals is considered to be a goodness-of-fit statistic, denoted as $X_d^2$, which is asymptotically chi-square distributed with $J-(q+1)$ degrees of freedom. The deviance residuals, the grouped deviance residuals, and the deviance goodness-of-fit statistic can be extracted from a fitted \textbf{RRglm} object with the commands, respectively:
\begin{verbatim}
residuals(out, type = "deviance")
residuals(out, type = "deviance.grouped")
RRglmGOF(object, doDeviance = TRUE, rm.na = TRUE)
\end{verbatim}
In the output of \textbf{RRglmGOF} the goodness-of-fit test(s) are reported, the p-value(s), the degrees of freedom, and the number of groups. For the H-L test, the results are also given for each cluster. The deviance statistic is equivalent to the LR statistic for testing the fitted model against the saturated model, which has a deviance of zero. The Pearson goodness-of-fit statistic is a score test statistic, also testing the fitted model against the saturated model. Thus, the Pearson and the deviance statistic are the score test and LR test for GLMs, respectively.

\paragraph{GLMM}
Different residuals can be considered for GLMMs. The response residual conditional on the random effect, is defined as the difference between the observation and the conditional expected value:
\begin{eqnarray}
r_{c}(y_{ik},\hat{\pi}_{ik}) &=& y_{ik} - E(y_{ik}\mid \mathbf{x}_{ik},\mathbf{z}_{ik},\mathbf{b}_{i}) \\
							 &=& y_{ik} - \pi\left(\mathbf{x}_{ik}^{t}\hat{\bm{\beta}} + \mathbf{z}_{ik}^{t}\hat{\mathbf{b}}_{i}\right) \nonumber
\end{eqnarray}
It can be extracted from the \textbf{residual} function by providing the argument \textbf{type="response"}. The unconditional response can be computed by integrating out the random effects to obtain the difference between the observation and the marginal mean. This residual is computed with the argument \textbf{type="unconditional.response"}.

The conditional and unconditional response residuals can be standardized by dividing them by their standard deviation, which leads to Pearson residuals:
\begin{verbatim}
residuals(object, type = "pearson")
residuals(object, type = "unconditional.pearson")
\end{verbatim}
Finally, as a result of sustained compatibility with \textit{lme4}, the usage of residuals aimed at \textbf{"merMod"} objects, such as working residuals and conditional deviance residuals, is not limited. Pearson residuals can also be scaled by any given user-specified weights.

\section{Software}\label{software}
The package \textit{GLMMRR} contains two main functions, \textbf{RRglm} and \textbf{RRglmer} for fitting a GLM and a GLMM given RR data, respectively. Both functions include the four link functions (logit, probit, cloglog, cauchit) for the different RR designs. The function \textbf{RRglm} makes a call to the function \textbf{glm} with the appropriate link function to fit a GLM for (binary) RR data. In the same way, the function \textbf{RRglmer} makes a call to the function \textbf{glmer} to fit a GLMM for (binary) RR data. The fit of both models, GLM and GLMM, is arranged by the computational routines of \textbf{glm} and \textbf{glmer}. Their general control parameters, the model and data-checking options, the type of optimizer, number of iterations can be specified in the \textbf{RRglm} and \textbf{RRglmer} functions. Thus, the numerical optimization algorithm and its specification can be defined in a similar way as in the call to functions \textbf{glm} and \textbf{glmer}.

The function \textbf{RRglm} and \textbf{RRglmer} creates an object of class \textbf{"RRglm"} and \textbf{"RRglmer"}, respectively. The package's summary, print and plot function can be used to get estimation results from an object of each class. Data needs to be defined in long format, where an RR specification is needed for each single case including the type of RR model and the design parameters. The package allows for different RR models and different design parameters across data cases. Together with a binary RR (outcome) variable and possible predictors, a GLM can be fitted. When also including a (factor) cluster variable, implying a correlation among clustered observations, a GLMM can be fitted. The complete functionality of the package can be accessed by making further input specifications.

\subsubsection{Input}
\begin{itemize}
\item \textbf{RRglm} \\
The general call to the function is: \textbf{RRglm(formula, link, item, RRmodel, p1, p2, data, na.action = "na.omit", ...)}
\begin{itemize}
\item formula: a two-sided linear formula object describing the model to be fitted, with the response on the left of a $\sim$ operator and the terms, separated by + operators, on the right.
\item
link: a GLM RRlink function for binary outcomes. Must be a function name; \textbf{RRlink.logit}, \textbf{RRlink.probit}, \textbf{RRlink.cloglog} and \textbf{RRlink.cauchit}.
\item item: optional item identifier, to obtain prevalence estimates per level of item.
\item RRmodel: the RR model per data case. Available options: \textbf{DQ}, \textbf{Warner}, \textbf{Forced}, \textbf{UQM}, \textbf{Crosswise}, \textbf{Triangular} and \textbf{Kuk}.
\item p1: the RR parameter $p_1$, defined per data case ($0 \le p_1 \le 1$).
\item p2: the RR parameter $p_2$, defined per data case ($0 \le p_2 \le 1$).
\item data: a data frame containing the variables named in formula as well as the RR model and parameters. If the required information cannot be found in the data frame, or if no data frame is given, then the variables are taken from the environment from which RRglm is called.
\item na.action: a function that indicates what should happen when the data contain NAs. The default action (na.omit, as given by getOption("na.action"))) strips any observations with any missing values in any variables.
\end{itemize}

\item \textbf{RRglmer}\\
The general call to the function is \textbf{RRglmer(formula, item, link, RRmodel, p1, p2, data, control = glmerControl(), na.action = "na.omit", ...)}
\begin{itemize}
\item formula: a two-sided linear formula object describing both the fixed and random effects part of the model, with the response on the left of a $\sim$ operator and the terms, separated by + operators, on the right. Random-effect terms are distinguished by vertical bars ("|") separating expressions for design matrices from grouping factors.
\item item	: optional item identifier to obtain prevalence estimates per level of item.
\item link: a GLM RRlink function for binary outcomes. Must be a function name; \textbf{RRlink.logit}, \textbf{RRlink.probit}, \textbf{RRlink.cloglog} and \textbf{RRlink.cauchit}.
\item RRmodel: the RR model per data case. Available options: \textbf{DQ}, \textbf{Warner}, \textbf{Forced}, \textbf{UQM}, \textbf{Crosswise}, \textbf{Triangular} and \textbf{Kuk}.
\item p1: the RR parameter $p_1$, defined per data case ($0 \le p_1 \le 1$).
\item p2: the RR parameter $p_2$, defined per data case ($0 \le p_2 \le 1$).
\item data: a data frame containing the variables named in formula as well as the RR model and parameters. If the required information cannot be found in the data frame, or if no data frame is given, then the variables are taken from the environment from which RRglmer is called.
\item na.action: a function that indicates what should happen when the data contain NAs. The default action (na.omit, as given by getOption("na.action"))) strips any observations with any missing values in any variables.
\item control: a list (of correct class, resulting from \textbf{lmerControl()} or \textbf{glmerControl()}, respectively, containing control parameters, including the nonlinear optimizer to be used and parameters to be passed through to the nonlinear optimizer, see the \textbf{lmerControl documentation} for details.
\end{itemize}
\end{itemize}

\subsubsection{Output}
\begin{itemize}
\item \textbf{RRglm}
An object of class \textbf{RRglm}, which extends the class \textbf{glm} with RR data. The object of class \textbf{RRglm} contains the regular GLM output and the following components:
\begin{itemize}
\item Item: the item levels for each data case -- prevalence rates are computed per level of item.
\item RRc: the RR-parameter $c$ for each data case (Equation (\ref{RRmodel})).
\item RRd: the RR-parameter $d$ for each data case (Equation (\ref{RRmodel})).
\item RRmodel: the RR model for each data case.
\item RRp1: the RR design parameter $p_1$.
 \item RRp2: the RR design parameter $p_2$.
\end{itemize}
\item \textbf{RRglmer}
An object of class \textbf{"RRglmerMod"}, which extends the class \textbf{"glmerMod"} with RR data. Many methods are available for the general class \textbf{"merMod"} to which
\textbf{"glmerMod"} and also the class \textbf{"RRglmerMod"} belongs. The methods for the \textbf{"merMod"}-class can be found in the documentation of \textit{lme4} and they are applicable to an object from class \textit{"RRglmerMod"}.
\end{itemize}

\section{Applications}\label{applications}

\subsection{An RR validation study}
\citet{hoglinger2018more} conducted an online experiment to validate different RRTs on the platform Amazon Mechanical Turk (\url{https://boris.unibe.ch/81516}). Participants were asked to play one of the two dice games, after which they were asked if they played honestly using randomly one of four RRTs. In the roll-a-six game the participant rolled a digital die by clicking a button, and was asked if the first roll resulted in a six. In the prediction game the participant was asked to think of a number, roll the digital die and was asked if the outcome corresponded to the memorized prediction. The participants were asked if they won in the dice game, and since both games relied on self-reports about the dice-roll outcomes cheating was easily possible. However, for the roll-a-six game the virtual dice outcomes were registered, and it was evaluated if participants illegitimately claimed a \$2 bonus payment. This enabled classifying the participants as \textbf{cheater} or \textbf{honest} (non-cheater) players. In the prediction game, individual cheating was not detectable. However, it was expected that around one sixth of all predictions were correct, since the dice outcomes were random. A systematic deviation from this percentage was attributed to cheating.

To disguise the true purpose of the study the survey was posted as an survey on “Mood and Personality” and included a range of questions, for instance questions on the big five personality trades. Participants were randomly appointed to one of four questioning methods (DQ, CW, FR, UQ), and were asked four sensitive questions. Besides the question about honest playing in the dice game, questions were asked about shoplifting, tax evasion, and voting. Although the answers to these last three questions could not be validated, the prevalence estimates of those who were identified as cheaters and non-cheaters in the dice games can be compared across RRTs.

\citet{hoglinger2018more} compared prevalence estimates of cheating in both games to those computed from the answers to the question whether they played honestly using the four RR techniques. They concluded that two RR techniques (FR, UQ) performed similar to DQ and did not reduce the level of misreporting. The level of underreporting was reduced by CW. However, CW also increased the level of overreporting, and the corresponding prevalence estimates were substantially higher than the true prevalence estimates of cheating.

The \textit{GLMMRR} is used to do a joint regression analysis of the RR data of the four questioning techniques. In a joint analysis the levels of under- and overreporting across RRTs can be directly compared through an interaction analysis of identified cheaters and the RRTs. The joint analysis is needed to quantify the different levels of underreporting across RRTs, and to identify who were misreporting. It is also examined if any background variables explain differences in misreporting across RRTs and sensitive questions.

The experiment had a two by three by five factorial design. Factor one represented the type of dice game (prediction game, roll-a-six game). Factor two was the type of RRT to ask the sensitive questions (DQ, CW, FR, UQ). The third factor was included to examine whether the implementations of the random devices produced the expected outcome distributions. This third factor is integrated in the current study by defining RRTs with different RR design parameters. \citet{hoglinger2018more} discuss the RRTs (factor 2) and the  random devices (factor 3) and the corresponding RR design parameters. Their supplement \textit{Documentation and codebook of the survey} was used to prepare the raw data for this study for analysis in R, which includes the specification of the design parameters for the RRTs. Our supplementary R script \textit{ASQ-MTurk data.R} comprises the code for preparing the data MTURK. The package \textit{GLMMRR} contains the prepared data object MTURK, which is constructed from the raw data using the code of the supplementary R script.

The true prevalence estimates of cheating are estimated for the roll-a-six game (\textbf{dicegame=2}) for the different RRTs. The estimates are based on the discrepancy between the actual dice outcome and the participant's response to whether the first roll was a six, which was asked through direct questioning. The true prevalence estimates are reported for each of the groups assigned to one of the RRTs.
\begin{verbatim}
R> package(GLMMRR)
R> data("MTURK", package="GLMMRR")
R> by(MTURK$cheaterdc[MTURK$dicegame==2],
+   MTURK$RRmodel[MTURK$dicegame==2],mean,na.rm=TRUE)

MTURK$RRmodel[MTURK$dicegame == 2]: DQ
[1] 0.04450262
------------------------------------------------------------------------------
MTURK$RRmodel[MTURK$dicegame == 2]: Crosswise
[1] 0.06032787
------------------------------------------------------------------------------
MTURK$RRmodel[MTURK$dicegame == 2]: UQM
[1] 0.05004812
------------------------------------------------------------------------------
MTURK$RRmodel[MTURK$dicegame == 2]: Forced
[1] 0.05188067
\end{verbatim}
Around 5\% of the participants cheated in the roll-a-six game and the estimates are comparable across RRT groups. It is examined if the prevalence estimates can also be recovered from the participant's responses to the question if they honestly reported whether a six was rolled (item \textbf{honest dice game reporting}, where response is coded as honest=0 and dishonest=1). This question was asked to all participants but different RRTs were used to obtain the response. The object is to validate the RRTs by comparing the prevalence estimate of cheating with the true prevalence estimate. The prevalence estimates are computed using the \textbf{RRglm} function where the \textbf{item} is the sensitive question if they reported honestly. The design parameters are stored in the data variables \textbf{RRp1} and \textbf{RRp2}, which are also reported in the output. It can be seen that for instance the CW design has two sets of parameters, where some participants were questioned with the CW method and design parameters .16 and 0, and others with .20 and 0.
\begin{verbatim}
R> MTURK_cheating1 <- MTURK[which(MTURK$dicegame==2 &
+   MTURK$Question=="cheating dice game"),]
R> rolla6 <- RRglm(RR_response ~ 1, item = Question,
+   link = "RRlink.logit", RRmodel = RRmodel,
+   p1=RRp1,p2=RRp2,data = MTURK_cheating1)
R> summary(rolla6)

### GLMMRR - Binary Randomized Response Data ###
Generalized linear fixed-effects model

Family:			 binomial
Link function:		 RRlogit

---------------------------------------------------------
Item:			 cheating dice game
Model(s):		 DQ (1.00 | 0.00)
			 Crosswise (0.16 | 0.00) (0.20 | 0.00)
			 UQM (0.78 | 0.49) (0.78 | 0.52)
			 Forced (0.75 | 0.67)

## Estimated Population Prevalence (weighted per RR model)
   RRmodel estimate.weighted se.weighted    n
 Crosswise          0.143393   0.0204156 1142
        DQ          0.039370   0.0099632  381
    Forced         -0.019361   0.0172690  769
       UQM          0.053496   0.0165870  778
\end{verbatim}
For the question \textit{honest dice game reporting} the estimated population proportion of dishonest reporters (ML estimate with standard errors) is reported for each RR design. Each RRT estimate is  the (weighted) average across different design parameters. It follows that the prevalence estimate of the CW method overestimates the true value of 6\% of detected cheaters who were CW questioned. Furthermore, the DQ and UQ perform approximately similar. The estimate is even negative for the FR technique. This can occur when some participants did not follow the RR instructions and/or when the random device distribution deviates from the expected distribution. This could also be an underlying problem of the CW method. A total of 33.6\% admitted that they do not know exactly the birthday of their parents. This could bias their response to the unrelated question which stated if their father/mother's birthday was in January or February (with expected probability 15.9\%), or between the 1st and the 6th of the month (with expected probability 19.7\%). The overestimation of the CW method could be caused by honest reporters who incorrectly answered \textbf{No} to the unrelated question about the birthday of one of their parents.

Misreporting is investigated further by computing the reduction in misreporting for each RRT for detected cheaters (dishonest reporters) and non-cheaters (honest reporters). Therefore, a dummy coded variable is defined for the cheaters (\textbf{cheaterdc=1}) and for each of the RRTs (DQ, CW, UQ, FR). A logistic regression analysis is performed using the \textbf{RRglm} function for the RR data of item \textbf{honest dice game reporting} (variable \textbf{RR\_response}) in the roll-a-six game conditional on the RRT and cheater identifiers.
\begin{verbatim}
R> rolla6A <- RRglm(RR_response ~ 1 + cheaterdc + CW + UQ + FR +
+   cheaterdc*CW+cheaterdc*UQ, item = Question,
+   link = "RRlink.logit", RRmodel = RRmodel, p1=RRp1, p2=RRp2,
+   data = MTURK_cheating1, na.action = "na.omit")
R> summary(rolla6A)

Coefficients:
             Estimate Std. Error z value Pr(>|z|)
(Intercept)   -4.8807     0.6043  -8.077 6.63e-16 ***
cheaterdc      5.8302     0.8087   7.209 5.63e-13 ***
CW             2.8283     0.6375   4.437 9.14e-06 ***
UQ             1.3453     0.8399   1.602   0.1092
FR            -1.3582     0.6941  -1.957   0.0504 .
cheaterdc:CW  -3.6246     0.9143  -3.964 7.36e-05 ***
cheaterdc:UQ  -2.0967     1.0798  -1.942   0.0522 .
---
Signif. codes:  0 ‘***’ 0.001 ‘**’ 0.01 ‘*’ 0.05 ‘.’ 0.1 ‘ ’ 1

(Dispersion parameter for binomial family taken to be 1)

    Null deviance: 3988.2  on 3069  degrees of freedom
Residual deviance: 2631.5  on 3063  degrees of freedom
  (5 observations deleted due to missingness)
AIC: 2645.5

Number of Fisher Scoring iterations: 7
\end{verbatim}
The reference level is the DQ technique (factor \textbf{RRmodel}) for those who were identified as honest reporters (factor \textbf{cheaterdc}) responding to being dishonest in the roll-a-six game. For DQ, the odds for dishonest reporting for non-cheaters is around zero ($\exp(-4.88)=0.008$), and for cheaters around 2.58. Of the DQ participants, who were identified as cheaters, the probability of admitting to be dishonest about whether a six was rolled is around 72\%. Participants might have guessed that this type of misreporting could be easily detected, which led to a non-substantial level of misreporting under DQ. The odds for dishonest reporting for non-cheaters in the UQ and FR condition are also close to zero. The UQ and FR effects are also not significantly different from zero. For the non-cheaters, the response technique, DQ, UQ or FR, did not influence their response. However, in the CW condition the odds of dishonest reporting for non-cheaters is significant and around .13, with a probability of 11.3\% of dishonest reporting, while being identified to be honest reporters. This led to the overestimation of the true prevalence rate by CW.

The opposite occurred for the cheaters in the CW and UQ condition. It was expected that the cheaters would be honest about their dishonesty in the roll-a-six game without the risk of disclosure in the CW and UQ condition. Instead, for both techniques the admitted level of misreporting by cheaters decreased. For cheaters in the CW and UQ condition, the probability of being dishonest is around 53.8\% and 54.9\%, respectively, which is much less than the 72\% under DQ. \citet{hoglinger2018more} argued that some cheaters might have misused the RRT to answer untruthfully without risk of detection who would have felt compelled to answer truthfully in DQ. It turned out that it was not possible to estimate an interaction effect for cheaters in the FR condition.

The three other sensitive questions (voting, shoplifting, tax evasion) do not have the problem that participants might be suspicious of being disclosed for their dishonesty as in the roll-a-six game. However, it is expected that the prevalence rates will be lower for the DQ group. Their responses are not masked and those participants are tended to underreport the questioned behavior in comparison to those not under risk of disclosure through the RR questioning techniques. Differences in prevalence rates across sensitive items (factor \textbf{Question}) and question techniques (factor \textbf{RRmodel}) are explored through a logistic regression using the \textbf{RRglm} function.
(factor \textbf{Question})
\begin{verbatim}
R> MTURK_cheating2 <- MTURK[which(MTURK$dicegame==2 &
+   MTURK$Question!="cheating dice game"),]
R> rolla6B <- RRglm(RR_response ~ 1 + RRmodel + Question +
+   cheaterdc*RRmodel, item = Question, link = "RRlink.logit",
+   RRmodel = RRmodel,p1=RRp1,p2=RRp2,data = MTURK_cheating2)

Coefficients:
                           Estimate Std. Error z value Pr(>|z|)
(Intercept)                -0.87793    0.08118 -10.815  < 2e-16 ***
RRmodelCrosswise            0.26976    0.09415   2.865  0.00417 **
RRmodelUQM                  0.39283    0.09279   4.233  2.3e-05 ***
RRmodelForced               0.11554    0.09783   1.181  0.23762
Questionshoplifting         0.61299    0.07040   8.708  < 2e-16 ***
Questiontax evasion        -1.01421    0.09100 -11.145  < 2e-16 ***
cheaterdc                   0.23951    0.31801   0.753  0.45135
RRmodelCrosswise:cheaterdc -0.04895    0.40461  -0.121  0.90371
RRmodelUQM:cheaterdc       -0.30170    0.42096  -0.717  0.47355
RRmodelForced:cheaterdc     0.32900    0.42022   0.783  0.43367
---
Signif. codes:  0 ‘***’ 0.001 ‘**’ 0.01 ‘*’ 0.05 ‘.’ 0.1 ‘ ’ 1

(Dispersion parameter for binomial family taken to be 1)

    Null deviance: 12705  on 9207  degrees of freedom
Residual deviance: 11797  on 9198  degrees of freedom
  (21 observations deleted due to missingness)
AIC: 11817

Number of Fisher Scoring iterations: 4
\end{verbatim}
The intercept corresponds to non-cheaters in the DQ condition responding to the item non-voting. For DQ, the odds for non-voting of the non-cheaters is around .41 ($\exp(-0.88)=0.41$), and for cheaters around 0.53. However, the cheaters did not respond significantly different from the non-cheaters as they did in reporting about being dishonest in the dice game. The prevalence rates of those questioned with a privacy protection (CW,UQ,FR) are higher -- the CW and UQ rates are significantly higher -- than of those questioned directly. The corresponding odds ratios are .54, 62, and .47 for CW, UQ, and FR, respectively. The prevalence rates for shoplifting are significantly higher and for tax evasion significantly lower than for non-voting. It is apparent that cheaters do not report significantly different under a privacy-protected response technique in comparison to the non-cheaters, since the interaction effects are approximately zero and non-significant. This occurred when cheaters were asked about their dishonesty in the roll-a-six game. This was probably provoked by the knowledge that their dishonesty could be detected under DQ. For the questions about non-voting, shoplifting, and tax evasion it was known that this was not possible, so cheaters were not/less inclined to misreport in the non-DQ condition.

The weighted prevalence rate (averaged over results from the same questioning technique with different design parameters) for each item and questioning technique are reported in the output. It follows that the prevalence rates are lower for DQ than for the other RRTs, but differences are small. Note that the reported significance in rates between CW and DQ and between UQ and DQ were averaged across the three items.
\begin{verbatim}
### GLMMRR - Binary Randomized Response Data ###
Generalized linear fixed-effects model

Family:			 binomial
Link function:		 RRlogit

---------------------------------------------------------
Item:			 non voting
Model(s):		 DQ (1.00 | 0.00)
			 Crosswise (0.16 | 0.00) (0.20 | 0.00)
			 UQM (0.78 | 0.49) (0.78 | 0.52)
			 Forced (0.75 | 0.67)

## Estimated Population Prevalence (weighted per RR model)
   RRmodel estimate.weighted se.weighted    n
 Crosswise           0.38215    0.022718 1142
        DQ           0.30607    0.023673  379
    Forced           0.33333    0.023720  768
       UQM           0.33889    0.022338  776

---------------------------------------------------------
Item:			 shoplifting
Model(s):		 DQ (1.00 | 0.00)
			 Crosswise (0.16 | 0.00) (0.20 | 0.00)
			 UQM (0.78 | 0.49) (0.78 | 0.52)
			 Forced (0.75 | 0.67)

## Estimated Population Prevalence (weighted per RR model)
   RRmodel estimate.weighted se.weighted    n
 Crosswise           0.46339    0.022914 1145
        DQ           0.44357    0.025452  381
    Forced           0.47479    0.024016  769
       UQM           0.56174    0.022936  778

---------------------------------------------------------
Item:			 tax evasion
Model(s):		 DQ (1.00 | 0.00)
			 Crosswise (0.16 | 0.00) (0.20 | 0.00)
			 UQM (0.78 | 0.49) (0.78 | 0.52)
			 Forced (0.75 | 0.67)

## Estimated Population Prevalence (weighted per RR model)
   RRmodel estimate.weighted se.weighted    n
 Crosswise           0.18850    0.021098 1143
        DQ           0.11549    0.016374  381
    Forced           0.13057    0.021182  771
       UQM           0.19407    0.020301  775
\end{verbatim}

The inclusion of the interaction effect between cheating status and RRT can be tested through a model comparison. The model is fitted without the interaction term (object \textbf{rolla6C}) and the \textbf{anova} function is used to produce a deviance table for the fitted objects. It follows that the interaction term does not lead to a significant model improvement.
\begin{verbatim}
R> anova(rolla6C, rolla6B,test="Chisq")
Analysis of Deviance Table

Model 1: RR_response ~ 1 + RRmodel + Question
Model 2: RR_response ~ 1 + RRmodel + Question + cheaterdc * RRmodel
  Resid. Df Resid. Dev Df Deviance Pr(>Chi)
1      9202      11802
2      9198      11797  4   5.3509   0.2532
\end{verbatim}

\subsubsection{Residual analysis}

The fit of the GLM with factor variables \textbf{RRT} and \textbf{Question} (object \textbf{rolla6C}) is evaluated with a residual analysis. The function \textbf{RRglmGOF} is used to compute overall goodness-of-fit tests. The Pearson and deviance chi-square statistics are computed given a grouping of the observations, based on the predictor variables. The H-L statistic is computed for ten groups of (approximately) equal size. There is evidence for a lack-of-fit when the statistic values are large.
\begin{verbatim}
R> RRglmGOF(RRglmOutput = rolla6C, doPearson = TRUE, doDeviance = TRUE,
+   doHlemeshow = TRUE)

GLMMRR - Binary Randomized Response Data

Goodness-of-Fit Testing
Response variable:		 RR_response
Predictor(s):			 RRmodel Question
Entries dataset:		 9208
---------------------------------------------------------
Summary:

                Statistic P.value df Groups
Pearson         12.95     0.0438  6  12
Deviance        13.01     0.0429  6  12
Hosmer-Lemeshow  8.82     0.3578  8  10

---------------------------------------------------------
\end{verbatim}
It follows that the Pearson and deviance goodness-of-fit statistics show a lack of fit, where the H-L statistic does not show a lack-of-fit. To further investigate the fit, the H-L statistic is computed for the same model but with different link functions. A different link function can improve the fit of the model and reduce the effects of a misspecified linear predictor. The estimated statistic values are given for the probit link function (see supplementary R-code for the complete analysis), which shows only a slight improvement in fit. The decrease in AIC is small and around .62.
\begin{verbatim}
R> rolla6E <- RRglm(RR_response ~ 1 + RRmodel + Question,
+   item = Question, link = "RRlink.probit",RRmodel = RRmodel,
+   p1=RRp1,p2=RRp2,data = MTURK_cheating2)
R> RRglmGOF(RRglmOutput = rolla6E, doPearson = TRUE, doDeviance = TRUE,
+   doHlemeshow = TRUE) ## improved fit over logit link

---------------------------------------------------------
Summary:

                Statistic P.value df Groups
Pearson         12.35     0.0545  6  12
Deviance        12.40     0.0537  6  12
Hosmer-Lemeshow  8.69     0.3689  8  10

---------------------------------------------------------
\end{verbatim}

\begin{figure}[h!]
    \centering
    \includegraphics{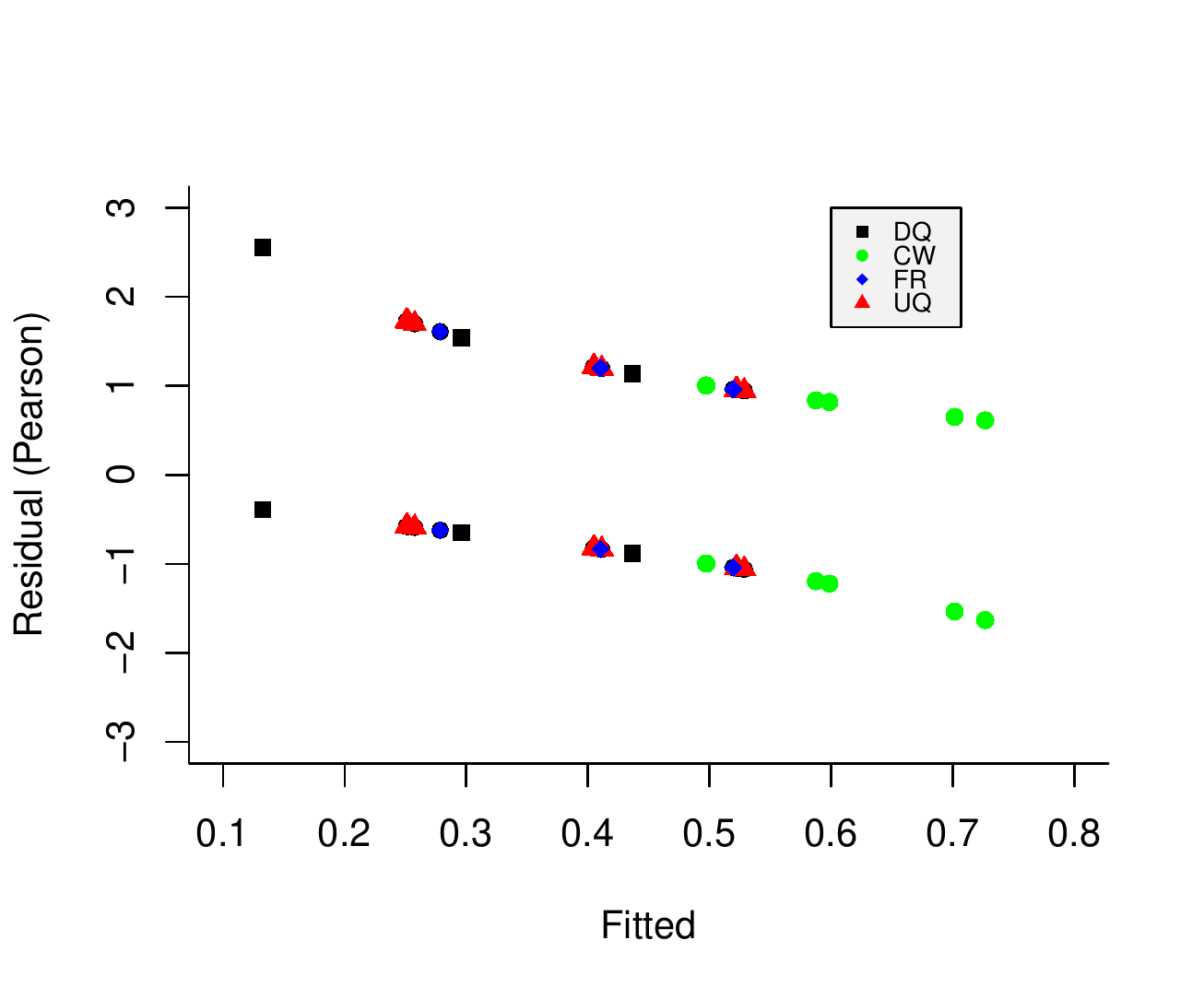}
     \caption{The roll-a-six game: for each RRT fitted success probabilities against Pearson residuals.}
     \label{fig:fitted-pearson}
  \end{figure}

Finally, the fitted probabilities are plotted against the estimated Pearson residuals for each RRT (R-code below) -- it is also possible to plot the object of class \textbf{"RRglm"}, \textbf{plot(rolla6C, which = 3, type = "pearson")}. In Figure \ref{fig:fitted-pearson},  the Pearson residuals (filled circles) under the CW method are large for the zero response observations. Typical for CW, high fitted probabilities (prevalence rates) correspond to zero and to one responses. There are three items, which leads to three different fitted probabilities for DQ. CW was used with two different sets of design parameters, which led to more than three fitted probabilities. The differences between prevalence rates and Pearson residuals for different design parameters is much larger for CW than for UQ and FR. For UQ and FR, the estimated Pearson residuals and fitted probabilities hardly differ across design parameters. It follows that the CW method is particularly sensitive to deviations from the design parameters. For an observed prevalence rate above .50, increasing the design parameter for CW will decrease the prevalence estimate. A possibility is that the design parameter for CW were incorrect and too low, which led to the overestimation of the prevalence rate for CW in the roll-a-six game. The residuals are also relatively large for low fitted probabilities, which is typical for DQ when the true prevalence rate is small. For reasons of brevity, we have omitted a further improvement of the model, and to explain individual differences in prevalence, by including individual predictor variables (e.g., gender, education, Big Five personality traits).

\begin{verbatim}
R> set <- names(rolla6C$linear.predictors)
R> dataset <- MTURK_cheating2[set,]
R> plot(rolla6C$fitted.values,residuals(rolla6C, type = "pearson"),
+   cex=.8,bty="l",xlim=c(.1,.8),ylim=c(-3,3),xlab="Fitted",
+   ylab="Residual (Pearson)")
R> points(rolla6C$fitted.values[which(dataset$DQ==1)],
+   residuals(rolla6C, type = "pearson")[which(dataset$DQ==1)],
+   xlim=c(0,1),ylim=c(-1,1),pch=15,col="black")
R> points(rolla6C$fitted.values[which(dataset$CW==1)],
+   residuals(rolla6C, type = "pearson")[which(dataset$CW==1)],
+   pch=16,col="green")
R> points(rolla6C$fitted.values[which(dataset$UQ==1)],
+   residuals(rolla6C, type = "pearson")[which(dataset$UQ==1)],
+   pch=17,col="red")
R> points(rolla6C$fitted.values[which(dataset$FR==1)],
+   residuals(rolla6C, type = "pearson")[which(dataset$FR==1)],
+   pch=18,col="blue")
R> legend(.6,3,c("DQ","CW","FR","UQ"),col=c("black","green",
+   "blue","red"),pch = c(15,16,18,17), bg = "gray95",cex=.7)
\end{verbatim}

\subsection{Extended item response modeling: Prevalence of student misconduct}

\citet{RePEcwpaper8} performed an online survey to estimate the prevalence of various forms of student misconduct (e.g., plagiarizing, cheating in exams). RRT was used, since students might be reluctant to reveal this kind of information. Four different types of RRTs were used (DQ, FR, CW, UQ). The RRTs were tailored to be implemented online which led to two implementations of FR (pick-a-numer, random wheel) and of CW (unrelated question, random wheel). In total there were six different RRTs to which participants were randomly assigned: DQ, two implementations of FR, UQ, and two implementations of CW. For all RRT implementations the level of protection was also varied, which led to different design probabilities. In total six different sets of parameters for UQ, two for FR, and eight for CW. The package \textit{GLMMRR} is used to do a joint regression analysis of the randomized item-response data, (1) to measure differences between RRTs, and (2) to examine differences in prevalence rates across conditions and students.

\begin{table}[h!]
\begin{tabular}{ccl}
\hline
 \textbf{No.}  && \multicolumn{1}{c}{\textbf{Item}}  \\ \cline{1-1} \cline{3-3}
1 && copied from other students during an exam (copied) \\
2 && used illicit crib notes in an exam (crib notes)  \\
3 && used prescription drugs to enhance your performance (drugs) \\
4 && handed in someone else’s work without citing (plagiarism)  \\
5 && had someone else write a large part of a submitted paper (someone else’s work) \\
\hline
\end{tabular}
\caption{\label{tab:tab1} Five items for measuring prevalence of student misconduct.}
\end{table}

Five sensitive items about student misconduct were surveyed using the six different RRTs with different design parameters. In Table \ref{tab:tab1}, the five items are given about respondent's own misconduct during exams and submitting a paper. The items were assumed to measure student misconduct. An RIRT model is used to measure each student's level of misconduct given RR data, while taking into account that students were assigned to different RRT conditions. A multiple-group (normal ogive) RIRT model is fitted, where the groups represent the RRT conditions. The function \textbf{RRglmer} of \textit{GLMMRR} is used with factor variable \textbf{Question} and factor variable \textbf{expcond} representing the items and experimental conditions (RRTs), respectively, for the randomized responses (\textbf{RR$\_$response}). The input \textbf{item} equals factor variable \textbf{Question} to obtain the (weighted) prevalence estimates for each item. The latent variable is represented by the person identifier \textbf{id}, where each person responded to the five items.

\begin{verbatim}
 R> out.re <- RRglmer(RR_response ~ 1 + Question + expcond + (1|id),
+   item=Question, link = "RRlink.probit", RRmodel = RRmodel,
+   p1=p1,p2=p2,data = ETHBE, control=glmerControl(
+   optimizer="bobyqa", optCtrl = list(maxfun = 200000)))
R> summary(out.re)

     AIC      BIC   logLik deviance df.resid
 21015.5  21103.2 -10496.7  20993.5    21394

Random effects:
 Groups Name        Variance Std.Dev.
 id     (Intercept) 0.4342   0.659
Number of obs: 21405, groups:  id, 4281

Fixed effects:
                                      Estimate Std. Error z value Pr(>|z|)
(Intercept)                           -1.20625    0.05633 -21.416  < 2e-16 ***
Questioncrib notes                    -0.34520    0.04582  -7.534 4.93e-14 ***
Questiondrugs                         -0.92016    0.05612 -16.395  < 2e-16 ***
Questionhanded in plagiarism          -0.83103    0.05369 -15.478  < 2e-16 ***
Questionhanded in someone else's work -1.06092    0.05982 -17.736  < 2e-16 ***
expcondFR pick-a-number                0.94880    0.06524  14.543  < 2e-16 ***
expcondCM pick-a-number                0.53036    0.08064   6.577 4.80e-11 ***
expcondFR random wheel                 0.99941    0.06520  15.329  < 2e-16 ***
expcondUQ Benford                      0.39748    0.07245   5.487 4.10e-08 ***
expcondCM unrelated question           0.67867    0.08017   8.465  < 2e-16 ***
---
Signif. codes:  0 ‘***’ 0.001 ‘**’ 0.01 ‘*’ 0.05 ‘.’ 0.1 ‘ ’ 1
\end{verbatim}
The levels of factor variable \textbf{Question} represents the item difficulties, where the intercept represents the first item (copied work). The levels of factor variable \textbf{expcond} represents the mean level of each RRT condition, and the intercept is the level of the DQ group.  The latent variable variance in student misconduct is around 0.43, which shows that, conditional on the item and experimental condition differences, around 30\% ($=.43/(1+.43)$) of the variance can be attributed to individual differences. The latent variable estimates represent the levels of student misconduct.

It is expected that students underreport their behavior. With an RRT individual responses are masked, and students are expected to answer more truthfully. Although students were randomly assigned to RRT conditions, the prevalence rates appear to be different across RRTs. The package \textit{multcomp} is used to test individual null hypotheses representing differences between  RRTs. Linear combinations of the experimental condition parameters are defined and tested simultaneously. Four hypothesis are evaluated, to test the difference in prevalence rates (1) between the two FR implementations (pick a number, random wheel), (2) between the two CW implementations (pick-a-number, unrelated question), (3) between FR and CW, (4) between FR and UQ. A matrix K is defined which represents the contrasts of interest, and the function \textbf{glht} is used to perform the general linear hypothesis testing.
\begin{verbatim}
R> K <- matrix(c(c(0, 0, 0, 0, 0, 1, 0, -1, 0 , 0),
+             c(0, 0, 0, 0, 0, 0, 1, 0, 0 ,-1),
+             c(0, 0, 0, 0, 0, 1, -1, 1, 0 ,-1),
+             c(0, 0, 0, 0, 0, 0, 0, 1, -1 ,0)), nrow=4,ncol=10,byrow=T)
R> t <- glht(out.re, linfct = K)
R> summary(t)

	 Simultaneous Tests for General Linear Hypotheses

Linear Hypotheses:
       Estimate Std. Error z value Pr(>|z|)
1 == 0 -0.05061    0.06176  -0.819    0.870
2 == 0 -0.14831    0.09125  -1.625    0.343
3 == 0  0.73917    0.11108   6.654   <0.001 ***
4 == 0  0.60193    0.06963   8.645   <0.001 ***
---
Signif. codes:  0 ‘***’ 0.001 ‘**’ 0.01 ‘*’ 0.05 ‘.’ 0.1 ‘ ’ 1
(Adjusted p values reported -- single-step method)
 \end{verbatim}
 It follows that the different implementations of FR does not lead to different prevalence rates. Both FR implementations provide a similar level of privacy protection. Furthermore, both CW implementations have similar prevalence rates. For FR and CW, differences between random devices do not lead to different prevalence estimates. The difference is significant between FR and CW, where FR produce higher rates than CW. FR also produced higher rates than UQ. It is remarkable that in this study students reported much higher rates with FR than CW, where in the RR validation study it was the other way around. It is not clear why with FR higher rates were obtained in the student misconduct study than with CW in comparison to the roll-a-six validation study. However, in the roll-a-six game, the prevalence rates with DQ also differed not much with those from FR.

\begin{figure}[h!]
    \centering
    \includegraphics[width=0.95\textwidth]{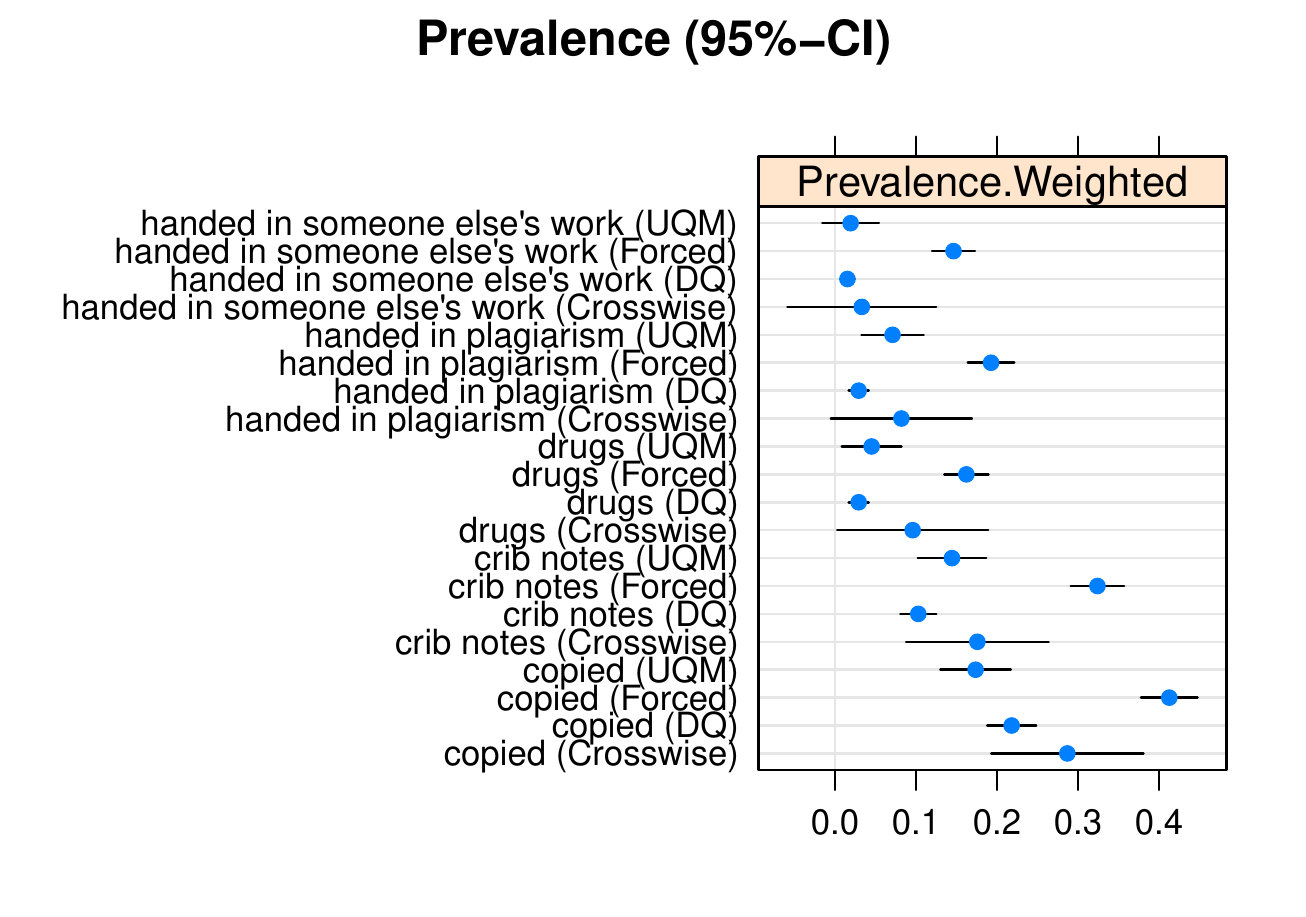}
     \caption{Prevalence estimates of types of student misconduct for different RRTs.}
     \label{fig:misconducts}
  \end{figure}

The prevalence rates for the five items are estimated for each RRT, averaging across results from different design parameters. In the output, an overview is given of the RRTs which were used to administer each item. For each RRT, the design parameters are given which were used. For instance, for item \textbf{copied}: DQ has one set of design parameters, UQ six sets of design parameters, FR two sets of design parameters, and CW eight sets of design parameters. Note that different design parameters were used to for different implementations of an online random device \citep{RePEcwpaper8}.
\begin{verbatim}
### GLMMRR - Binary Randomized Response Data ###
Generalized linear mixed-effects model

Family:			 binomial
Link function:		 RRprobit

---------------------------------------------------------
Item:			 copied
Model(s):		 DQ (1.00 | 0.00)
			 UQM (0.70 | 0.49) (0.70 | 0.50) (0.70 | 0.52)
			     (0.78 | 0.49) (0.78 | 0.50) (0.78 | 0.52)
			 Forced (0.67 | 0.17) (0.75 | 0.17)
			 Crosswise (0.17 | 0.00) (0.20 | 0.00) (0.23 | 0.00)
(0.25 | 0.00) (0.26 | 0.00) (0.30 | 0.00) (0.75 | 0.00) (0.83 | 0.00)

## Estimated Population Prevalence (weighted per RR model)
   RRmodel estimate.weighted se.weighted    n
 Crosswise           0.28679    0.047932 1427
        DQ           0.21806    0.015389  720
    Forced           0.41265    0.017797 1417
       UQM           0.17351    0.022187  717

---------------------------------------------------------
Item:			 crib notes
Model(s):		 DQ (1.00 | 0.00)
			 UQM (0.70 | 0.49) (0.70 | 0.50) (0.70 | 0.52)
			     (0.78 | 0.49) (0.78 | 0.50) (0.78 | 0.52)
			 Forced (0.67 | 0.17) (0.75 | 0.17)
			 Crosswise (0.17 | 0.00) (0.20 | 0.00) (0.23 | 0.00)
(0.25 | 0.00) (0.26 | 0.00) (0.30 | 0.00) (0.75 | 0.00) (0.83 | 0.00)

## Estimated Population Prevalence (weighted per RR model)
   RRmodel estimate.weighted se.weighted    n
 Crosswise           0.17570    0.044884 1427
        DQ           0.10278    0.011317  720
    Forced           0.32396    0.016822 1417
       UQM           0.14438    0.021528  717

---------------------------------------------------------
Item:			 drugs
Model(s):		 DQ (1.00 | 0.00)
			 UQM (0.70 | 0.49) (0.70 | 0.50) (0.70 | 0.52)
			     (0.78 | 0.49) (0.78 | 0.50) (0.78 | 0.52)
			 Forced (0.67 | 0.17) (0.75 | 0.17)
			 Crosswise (0.17 | 0.00) (0.20 | 0.00) (0.23 | 0.00)
(0.25 | 0.00) (0.26 | 0.00) (0.30 | 0.00) (0.75 | 0.00) (0.83 | 0.00)

## Estimated Population Prevalence (weighted per RR model)
   RRmodel estimate.weighted se.weighted    n
 Crosswise          0.095850   0.0475537 1427
        DQ          0.029167   0.0062712  720
    Forced          0.162119   0.0138931 1417
       UQM          0.045145   0.0187692  717

---------------------------------------------------------
Item:			 handed in plagiarism
Model(s):		 DQ (1.00 | 0.00)
			 UQM (0.70 | 0.49) (0.70 | 0.50) (0.70 | 0.52)
			     (0.78 | 0.49) (0.78 | 0.50) (0.78 | 0.52)
			 Forced (0.67 | 0.17) (0.75 | 0.17)
			 Crosswise (0.17 | 0.00) (0.20 | 0.00) (0.23 | 0.00)
(0.25 | 0.00) (0.26 | 0.00) (0.30 | 0.00) (0.75 | 0.00) (0.83 | 0.00)

## Estimated Population Prevalence (weighted per RR model)
   RRmodel estimate.weighted se.weighted    n
 Crosswise          0.081899   0.0444126 1427
        DQ          0.029167   0.0062712  720
    Forced          0.192582   0.0145974 1417
       UQM          0.070978   0.0195603  717

---------------------------------------------------------
Item:			 handed in someone else's work
Model(s):		 DQ (1.00 | 0.00)
			 UQM (0.70 | 0.49) (0.70 | 0.50) (0.70 | 0.52)
			     (0.78 | 0.49) (0.78 | 0.50) (0.78 | 0.52)
			 Forced (0.67 | 0.17) (0.75 | 0.17)
			 Crosswise (0.17 | 0.00) (0.20 | 0.00) (0.23 | 0.00)
 (0.25 | 0.00) (0.26 | 0.00) (0.30 | 0.00) (0.75 | 0.00) (0.83 | 0.00)

## Estimated Population Prevalence (weighted per RR model)
   RRmodel estimate.weighted se.weighted    n
 Crosswise          0.033014   0.0468658 1427
        DQ          0.015278   0.0045711  720
    Forced          0.146240   0.0134956 1417
       UQM          0.019240   0.0178254  717
\end{verbatim}
The prevalence estimates are almost always lowest for DQ. Most likely students underreported their behavior when directly asked. In Figure \ref{fig:misconducts}, the weighted prevalence estimates are plotted for each item and RRT. This is the first of four plots, when plotting an object of class \textbf{RRglmerMod} of the package \textit{GLMMRR}. It can be seen that copying work is the most popular way, and then using a crib note. Although the CW condition has the highest number of students, the confidence interval of the CW estimates are much wider than for the other conditions. The CW method is less efficient in estimating student prevalence than the other RRTs.

\subsubsection{Model fit}
The relevance of the random person component is examined by comparing the multiple-group IRT model with a (multiple-group) GLM, which has the same fixed effect part.
\begin{verbatim}
R> out.fe <- RRglm(RR_response ~ 1 + Question + expcond, item=Question,
+      link = "RRlink.probit",RRmodel = RRmodel,p1=p1,p2=p2,data = ETHBE)
R> anova(out.re,out.fe)
Data: df
Models:
out.fe: RR_response ~ 1 + Question + expcond
out.re: RR_response ~ 1 + Question + expcond + (1 | id)
       npar   AIC   BIC logLik deviance  Chisq Df Pr(>Chisq)
out.fe   10 21147 21227 -10564    21127
out.re   11 21016 21103 -10497    20994 133.82  1  < 2.2e-16 ***
\end{verbatim}
It follows that the RIRT model fits the data better than the GLM, with a lower AIC and BIC, and a significant decrease in deviance. For the RIRT model, conditional Pearson residuals are computed given the random effect. The Pearson residuals are plotted against the fitted probabilities. To obtain the fitted probabilities, the fitted values on the linear predictor scale are computed using the \textbf{predict} function. Then, according to Equation $\ref{RRmodel}$, the fitted probabilities are computed using parameters $c$ and $d$, which can computed from function \textbf{getRRparameters} given the RRT and RR design parameters.
\begin{verbatim}
R> eta <- predict(out.re, type = "link")
R> dum <- getRRparameters(ETHBE$RRmodel, ETHBE$p1, ETHBE$p2)
R> pp <- dum$c + dum$d*pnorm(eta)
R> resid <- residuals(out.re, type = c("pearson"))
R> plot(pp[ETHBE$expcond=="direct questioning"],
+   resid[ETHBE$expcond=="direct questioning"],bty="l",
+   xlim=c(0,1),ylim=c(-2,6),cex=.8,xlab="Fitted probability",ylab="Pearson residual")
R> points(pp[ETHBE$expcond=="CM pick-a-number"],
+   resid[ETHBE$expcond=="CM pick-a-number"],cex=.8,pch=19,col="grey80")
R> points(pp[ETHBE$expcond=="FR pick-a-number"],
+   resid[ETHBE$expcond=="FR pick-a-number"],cex=.8,pch=15,col="red")
R> points(pp[ETHBE$expcond=="FR random wheel"],
+   resid[ETHBE$expcond=="FR random wheel"],cex=.8,pch=16,col="blue")
R> points(pp[ETHBE$expcond=="CM unrelated question"],
    resid[ETHBE$expcond=="CM unrelated question"],cex=.8,pch=17,col="green")
R> points(pp[ETHBE$expcond=="UQ Benford"],
+    resid[ETHBE$expcond=="UQ Benford"],cex=.8,pch=18,col="purple")
R> abline(h=1,lty=2,col="grey")
R> abline(h=-1,lty=2,col="grey")
R> legend(.6,6,c("DQ","FR pick-a-number","FR random wheel",
+   "CM unrelated", "UQ","CM pick-a-number"),
+   col=c("black","red","blue","green","purple","grey80"),
+   pch = c(1,15,16,17,18,19),cex=.7,bty="n")
\end{verbatim}

\begin{figure}[h!]
    \centering
    \includegraphics[width=0.95\textwidth]{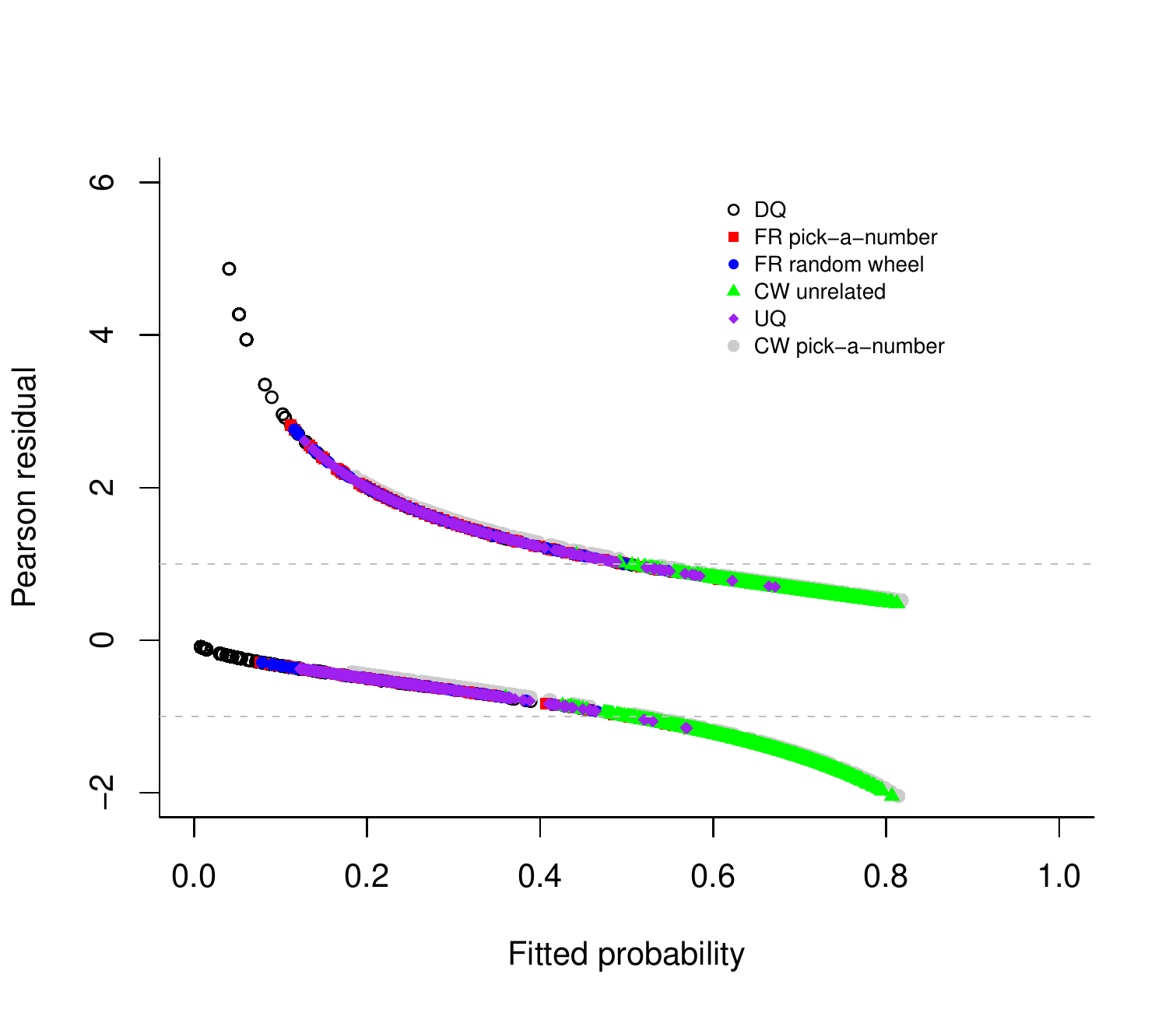}
     \caption{Fitted probabilities of types of student misconduct for different RRTs versus the Pearson residuals.}
     \label{fig:residprevalence}
  \end{figure}

In Figure \ref{fig:residprevalence}, the Pearson residuals are plotted for each RRT. It can be seen that for DQ the residuals are large (small) for positive (zero) responses, since the prevalence rates are small. The residuals for FR and UQ are relatively small, partly because the corresponding fitted probabilities are in the middle of the scale. For the CW methods,  it can be seen that the residuals are large (small) for zero (positive) responses, since they correspond to high prevalence rates. A further analysis to improve the model by incorporating predictor variables is omitted for reasons of brevity.

\section{Discussion}\label{conclusion}

The \textit{GLMMRR} provides tools for fitting GLM(M)s on RR data, where the RR design and the design parameters are allowed to vary across observations. The multiple-group RR designs are particularly interesting to validate RR methods, to examine the sensitivity of the attributes, and to examine the influence of different levels of privacy protection. The multiple group modelling approach also supports simultaneously testing RR design effects across items and participants. The tools provide support to substantive applications using RR techniques, which can include different RR designs and advanced GLM and GLMM methods to analyse RR data. It is our objective to stimulate applied and methodological RR research by offering the
open-source software \textit{GLMMRR}. The tools extend the popular modeling tools of \textit{lme4}, and class functions are simply extended to deal with RR data, while maintaining the general features included in the GLM and GLMM software \citep[e.g., \textit{lme4};][]{bates2015fitting}.

When some respondents do not follow the RR design instructions, the GLM(M) does not fit the data \citep[e.g.,][]{bockenholt2007item,fox2013mixture,doi:10.1509/jmkr.47.1.14}. In that case, a composite link function can be used to include a linear predictor for those not following the instructions and one for those following the instructions. Currently, the ML estimations methods for \textit{GLMMRR} cannot handle composite link functions. More research is needed to develop and implement estimation methods that can handle in a flexible way GLMMs with composite link functions \citep{thompson1981composite}.

The modified link functions can be used for other type of GLMMs. In longitudinal research, in practice, time is often observed in discrete units. The discrete-time hazard defines the probability of the occurrence of an event at time $t$. The GLM(M) can be used to describe the link between the hazard rate and a linear predictor. The GLM(M) with the complementary log-log link function applies, when the data are generated by a continuous-time proportional hazards model \citep{allison1982discrete}. When using an RR design to collect information about sensitive events (e.g., events related to war, trauma, sexual assault), the GLMM with a modified link function is an appropriate model to analyse the RR data. For count RR data, the complementary log-log link function can be used to model the probability of an RR of a non-zero observation \citep{fox2018generalized}. Ordinal data can be modeled with a (cumulative) link function for the proportional odds, which is defined by cumulative probabilities. For each response category, a GLMM defines a cumulative probability that a response falls below or in this category. The modified link functions can be used to model cumulative probabilities with GLMMs for ordinal RR data.

The GLMMs can be computationally intensive and usually require relatively large sample sizes. RR designs also require larger samples sizes to achieve the same level of accuracy as DQ. Furthermore, the prevalence of sensitive behaviors is often relatively low, and to obtain reliable estimates more data is required. The \textit{GLMMRR} package supports large sample sizes, which can be around $10^6$ observations.

\bibliographystyle{apacite}
\bibliography{refs-RR-update}

\end{document}